\documentclass{aastex}
\usepackage{spr-astr-addons}
\usepackage{mathptmx}

\RequirePackage{color}
\def \etal {{\it et al. \,}}

\begin{document}

\title{The Centurion 18 telescope of the Wise Observatory}
\shorttitle{The C18 at Wise} \shortauthors{Brosch et al.}

\author{Noah Brosch}
\affil{The Wise Observatory and the Raymond and Beverly Sackler
School of Physics and Astronomy, The Raymond and Beverly Sackler
Faculty of Exact Sciences, Tel Aviv University, Tel
Aviv 69978, Israel. Email: noah@wise.tau.ac.il} 
\author{David Polishook} \affil{The Wise Observatory and Department of Geophysics
and Planetary Sciences, The Raymond and Beverly Sackler Faculty of
Exact Sciences, Tel Aviv University, Tel
Aviv 69978, Israel. Email: david@wise.tau.ac.il} 
\author{Avi Shporer} \affil{The Wise Observatory and the Raymond and Beverly Sackler
School of Physics and Astronomy,  The Raymond and Beverly Sackler
Faculty of Exact Sciences, Tel Aviv University, Tel
Aviv 69978, Israel. Email: shporer@wise.tau.ac.il} 
\author{Shai Kaspi} \affil{The Wise Observatory and the Raymond and Beverly Sackler
School of Physics and Astronomy,  The Raymond and Beverly Sackler
Faculty of Exact Sciences, Tel Aviv University, Tel
Aviv 69978, Israel. Email: shai@wise.tau.ac.il} 
\author{Assaf Berwald} \affil{Astronomy manager of the NANA forum
(http://www.nana10.co.il).\.Email:\,assafberwald@gmail.com} 
\author{Ilan Manulis} \affil{Technoda Center for
Education in Science and Technology,  P.O. Box 1144 Givat Olga,
Hadera 38110, Israel. Email: ilan@trendline.co.il}



\begin{abstract}
We describe the second telescope of the Wise Observatory, a 0.46-m
Centurion 18 (C18) installed in 2005, which enhances significantly
the observing possibilities. The telescope operates from a small
dome and is equipped with a large-format CCD camera. In the last
two years this telescope was intensively used in a variety of
monitoring projects.

The operation of the C18 is now automatic, requiring only start-up
at the beginning of a night and close-down at dawn. The
observations are mostly performed remotely from the Tel Aviv
campus or even from the observer's home. The entire facility was
erected for a component cost of about 70k\$ and a labor investment
of a total of one man-year.

We describe three types of projects undertaken with this new
facility: the measurement of asteroid light variability with the
purpose of determining physical parameters and binarity, the
following-up of transiting extrasolar planets, and the study of
AGN variability. The successful implementation of the C18
demonstrates the viability of small telescopes in an age of huge
light-collectors, provided the operation of such facilities is
very efficient.

\end{abstract}


\section{Introduction}
\begin{figure*}
{\centering
 \includegraphics[clip=,angle=0,width=17cm]{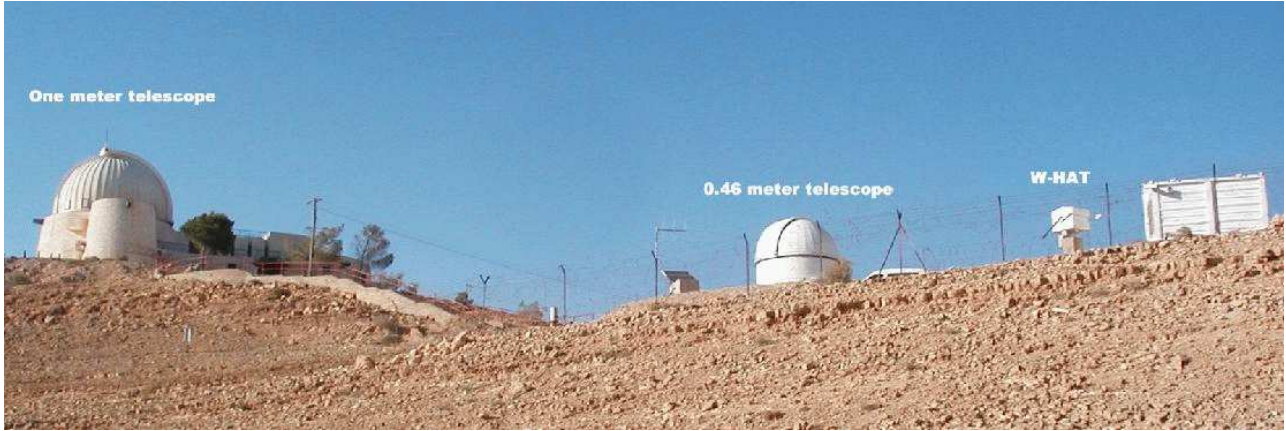}
}
 \caption{%
The Wise Observatory observing facilities. From left to right: the
dome of the T40 and main building, the dome of the C18, and the
box enclosure of the WHAT.
    \label{fig:WiseObs}}
\end{figure*}

The Wise Observatory (WO, see http://wise-obs.tau.ac.il) began
operating in 1971 as a Tel-Aviv University (TAU) research
laboratory in observational optical astronomy. It is located on a
high plateau in the central part of the Negev desert (longitude
34$^\circ$45' 48" E, latitude 30$^\circ$35'45" N, altitude 875 m,
time zone is -2 hours relative to Universal Time). The site is
about 5 km west of the town of Mitzpe Ramon, 200 km south of
Tel-Aviv and 86 km south of Beer Sheva. An image of the Wise
Observatory is shown in Figure~\ref{fig:WiseObs}.

The WO was originally equipped with a 40-inch telescope (T40). The
Boller and Chivens telescope is a wide-field Ritchey-Chr\'{e}tien
reflector mounted on a rigid, off-axis equatorial mount. The
optics are a Mount Wilson/Palomar Observatories design, consisting
of a 40-inch diameter clear aperture f/4 primary mirror, a
20.1-inch diameter f/7 Ritchey-Chr\'{e}tien secondary mirror, and
a quartz corrector lens located 4 inches below the surface of the
primary mirror, providing a flat focal field of up to 2.5 degrees
in diameter with a plate scale of 30 arcsec mm$^{-1}$. An f/13.5
Cassegrain secondary mirror is also available, but is hardly used
nowadays. This telescope was originally a twin of the Las Campanas
1m Swope telescope, described by Bowen \& Vaughan (1973), though
the two instruments diverged somewhat during the years due to
modifications and upgrades. The telescope is controlled by a
control system located in the telescope room.

In its 36 years of existence the observatory has kept abreast of
developments in the fields of detectors, data acquisition, and
data analysis. In many instances the observations can now be
performed remotely from Tel Aviv, freeing the observer from the
necessity to travel to the observing site. The efficiency of
modern detectors implies that almost every photon collected by the
telescope can be used for scientific analysis.

The on-going modernization process allowed landmark studies to be
performed and generations of students to be educated in the
intricacies of astronomy and astrophysics. Some of these students
are now staff members of the Physics and Astronomy Department at
TAU or at other academic institutions in Israel or overseas. The
one-meter telescope is over-subscribed, with applications for
observing time exceeding by $\sim$50\% the number of available
nights. This demonstrates the vitality of the observatory as a
research and academic education facility, even though on a world
scale the size of the telescope shrank from being a medium-sized
one in the early-1970s to being a ``small'' telescope nowadays.

The Wise Observatory operates from a unique location, in a time
zone between India and Greece and in a latitude range from the
Caucasus to South Africa, where no other modern observatories
exist, and at a desert site with a large fraction of clear nights.
Thus, even though its 1m telescope is considered small, it is
continuously producing invaluable data for the study of
time-variable phenomena, from meteors and extrasolar planet
searches to monitoring GRBs and microlensing events, to finding
distant supernovae and ``weighing'' black holes in AGNs. A cursory
literature search shows that the Wise Observatory has one of the
greatest scientific impacts among 1m-class telescopes.

One research aspect that developed into a major WO activity branch
is of time-series studies of astronomical phenomena. A project to
monitor photometrically and spectroscopically Active Galactic
Nuclei (AGNs) is still running, following about 30 years of data
collection. Other major projects include searches for supernovae
or for extrasolar planets (using transits or lensing events),
observations of novae and cataclysmic variables, studies of
star-forming galaxies in a variety of environments, and studies of
Near Earth Objects (NEOs) and other asteroids. These studies,
mainly part of PhD projects, are observation-intensive and require
guaranteed telescope access for a large number of nights and for a
number of years. The oversubscription of the available nights on
the T40, the need to follow-up possible discoveries by small
telescopes with long observing runs, and a desire to provide a
fallback capability in case of major technical problems with the
T40, required therefore the expansion of the WO observational
capabilities.

Small automatic, or even robotic, instruments provide nowadays
significant observing capability in many astronomy areas.
Combining these small telescopes into a larger network can
increase their scientific impact many-fold. This is currently done
with the GCN (Barthelmy et al. 1998). Examples of such instruments
are REM (Zerbi \etal 2001), WASP and super-WASP (e.g., Pollacco
\etal 2006), HAT (Bakos \etal 2004), and ROTSE-III (e.g., Akerlof
\etal 2003; Yost 2006), and very recently KELT (Pepper \etal
2007). Among the larger automatic instruments we mention the
Liverpool Telescope (e.g., Steele \etal 2004) and its clones. Such
automatic instruments enable the exploration of the last poorly
studied field of astronomy: the temporal domain. However, a first
step before developing a network of automatic telescopes is
demonstrating feasibility at a reasonable cost for a single, first
instrument. This paper describes such an experiment at the Wise
Observatory in Israel.

\section{The telescope}
\label{txt:Telescope}

To enhance the existing facilities with special emphasis on
time-series astronomy, we decided in 2002 to base an additional
observing facility on as many off-the-shelf hardware and software
components as possible, to speed the development and bring the new
facility on-line as soon as possible. We decided to acquire a
Centurion 18 (C18) telescope manufactured by AstroWorks, USA. The
telescope was delivered towards the end of 2003 and operated for a
year in a temporary enclosure. From 2005 the telescope operates in
its permanent dome, which is described below. The C18 has a
prime-focus design with an 18-inch (0.46-m) hyperbolic primary
mirror figured to provide an f/2.8 focus. The light-weighted
mirror reflects the incoming light to the focal plane through a
doublet corrector lens. The telescope is designed to image on
detectors as wide as regular (35-mm) camera film though with
significant edge-of-field vignetting, but we now use a much
smaller detector, and the images are very reasonable indeed
(FWHM=$\sim$2".9 at the edge of the CCD, only 13\% worse than at
field center, with most image size attributable to local seeing).

\begin{figure*}
{\centering
 \includegraphics[clip=,angle=0,width=17cm]{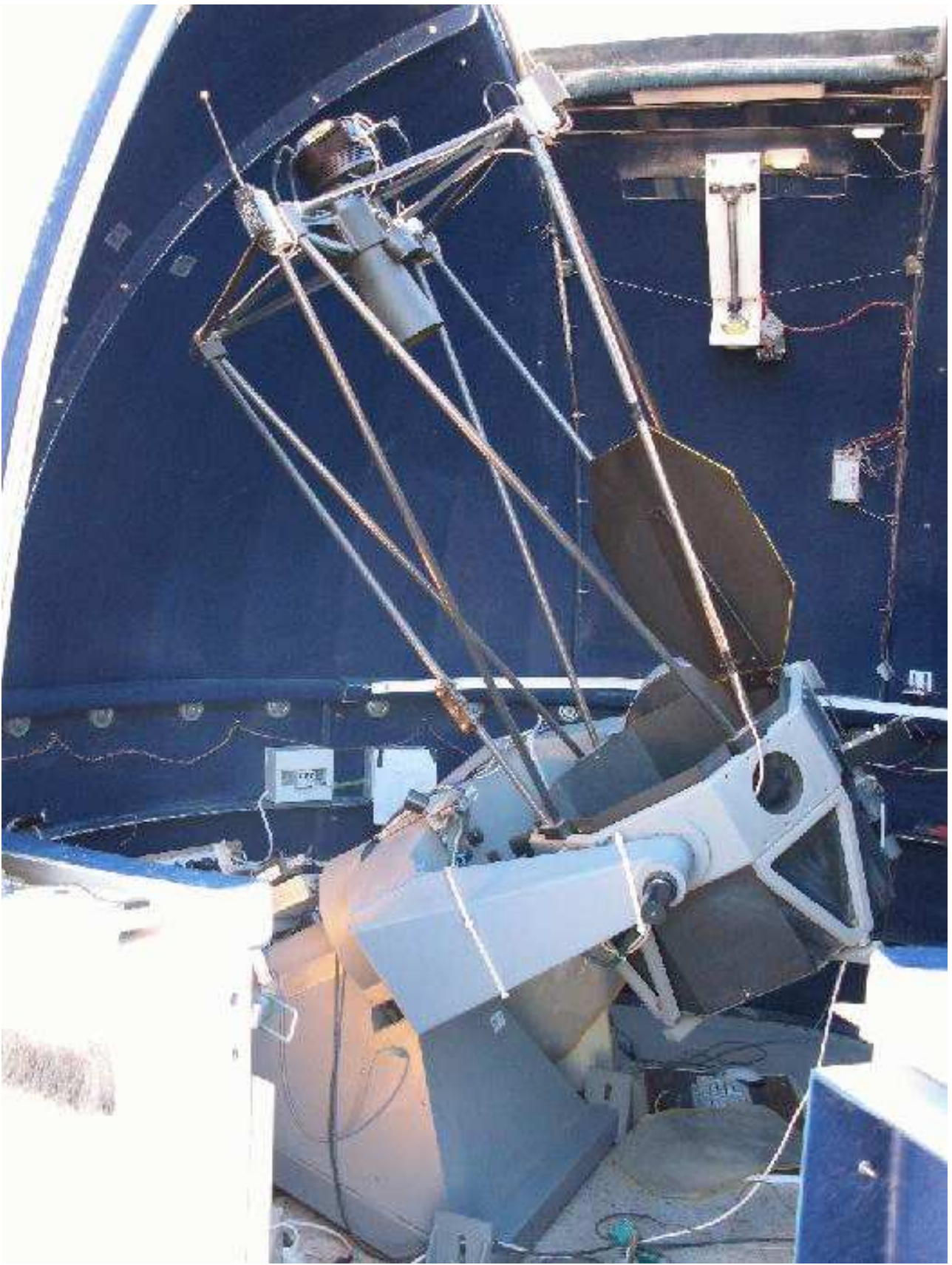}
}
 \caption{%
The Centurion 18 telescope viewed through the dome slit, with the
dome door opened. Note the anti-collision switches mounted at the
top of the truss structure and on the equatorial fork, and the
electrically-operated primary mirror cover, here in the open
position.
    \label{fig:C18_in_dome}}
\end{figure*}

The focal plane is maintained at the proper distance from the
primary mirror by a carbon-reinforced epoxy plastic (CREP) truss
tube structure. The support tubes allow the routing of various
wires and tubes (see below) through the structure, providing a
neat construction and much less vignetting than otherwise. The
CREP structure has a very low expansion coefficient; this implies
that the distance between the primary and the focal plane hardly
changes with temperature. The stiff structure and the relatively
low loading of the focal plane imply that the telescope hardly
flexes with elevation angle.

The optical assembly uses a fork mount, with the right ascension
(RA) and declination (DEC) aluminum disk drives being of
pressure-roller types. The steel rollers are rotated by stepping
motors and both axes are equipped with optical sensors, thus the
motions of the telescope can be controlled by computer. The fork
mount and the truss structure limit the telescope pointing to
north of declination -33$^{\circ}$.

The focal plane assembly permits fine focusing using a
computer-controlled focuser connected to the doublet corrector and
equipped with a stepper motor. The C18 was originally supplied
with a primary mirror metal cover that had to be removed manually.
Since then, an electrically-operated, remotely-commanded mirror
cover was installed.

\section{The CCD}
\label{txt:CCD} The telescope was equipped from the outset with a
Santa Barbara Instrument Group (SBIG) ST-10 XME USB CCD camera
that was custom-fitted to our specific telescope by AstroWorks and
was delivered together with the telescope. This
thermoelectrically-cooled chip has 2184$\times$1472 pixels each
6.8 $\mu$m wide, which convert to 1.1 arcsec at the f/2.8 focus of
the telescope. The chip offers, therefore, a 40'.5$\times$27'.3
field of view. A second, smaller CCD, mounted next to the science
CCD, allows guiding on a nearby star using exactly the same
optical assembly. The CCD is used in ``white light'' with no
filter.
The readout noise is 10 electrons per pixel and
the gain is 1.37 electrons per count. Each FITS image is 6.4 MB
and the read out time is $\sim$15 sec using the {\it MaximDL}
package.

The CCD is mounted behind the doublet corrector lens and the
focusing is achieved by moving the lens. In practice, even though
the beam from the telescope is strongly converging, the focus is
fairly stable throughout the night despite ambient temperature
excursions of $\sim10^{\circ}$C or more. In any case, refocusing
is very easy using the automatic focuser.

After a few months of test operations, we decided to add water
cooling to the ST-10. This was achieved by feeding the inlet water
port of the SBIG CCD with an antifreeze solution from a one-liter
reservoir that was continuously circulated by an aquarium pump.
The addition of water cooling reduced the CCD temperature by an
additional 10$^{\circ}$C, lowering the dark counts to below 0.5
electrons pixel$^{-1}$ sec$^{-1}$. We emphasize that the water
circulation operates continuously, 24 hours per day, irrespective
of whether the telescope is used for observations or not. Since
its implementation, we changed the cooling fluid only once when it
seemed to develop some kind of
plaque.

The CCD is used in ``white light'' without filters to allow the
highest possible sensitivity. The chip response reaches 87\%
quantum efficiency near 630 nm, implying an overall response
similar to a ``wide-R'' band. 

\begin{figure*}
{\centering
 \includegraphics[clip=,angle=-90,width=8.5cm]{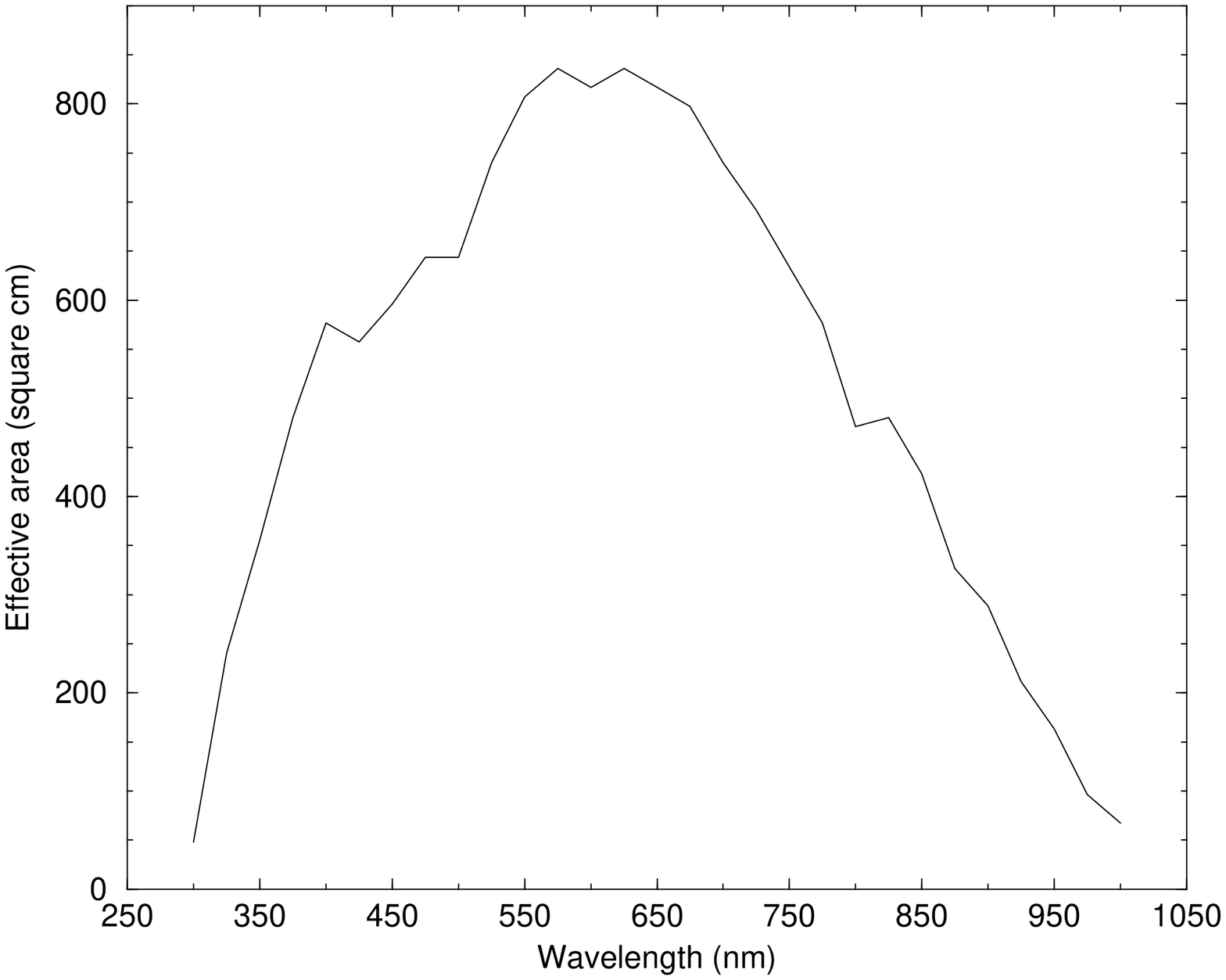}
}
 \caption{%
Effective area of the C18 telescope with the SBIG ST-10XME CCD.
    \label{fig:C18_response}}
\end{figure*}

The effective area of the telescope and CCD is shown in
Figure~\ref{fig:C18_response}. This was calculated using the
generic CCD response from the SBIG web site for the Kodak enhanced
KAF-3200ME chip, assuming 7\% areal obscuration of the primary
mirror by the secondary mirror baffle and CCD, 80\% mirror
reflectivity, and 5\% attenuation at each surface by the doublet
lens and CCD window.

\section{The dome}
\label{txt:Dome}

Given the small size of the telescope and the high degree of
automation desired, we chose a small dome that would not allow
routine operation with a human inside, but would allow
unrestricted access to the sky for the C18. From among the
off-the-shelf domes we chose a Prodome 10-foot dome from Technical
Innovations USA\footnote{http://www.homedome.com/}, equipped with
two wall rings to provide sufficient height for the C18 in all
directions.

The fiberglass dome was equipped by Technical Innovations with an
electrical shutter and with the necessary sensors to operate the
{\it Digital Dome Works (DDW)} software bundled with the dome;
this allows control of all the dome functions from a computer
equipped with a four-port Multi-I/O card, adding four serial ports
to the two already available on the operating computer. In
addition, we equipped the dome with a weather station mounted on a
nearby mast, a video camera with a commandable light source that
allows remote viewing of the telescope and of part of the dome to
derive indications about its position, and a ``Robo reboot''
device for {\it DDW}. The latter was installed to allow the remote
initialization of the dome functions in case of a power failure.

Later, we added a Boltwood Cloud
Sensor\footnote{http://www.cyanogen.com/products/cloud\_main.htm}
that measures the amount of cloud cover by comparing the
temperature of the sky to that of the ambient ground. The sky and
ground temperatures are determined by measuring the amount of
radiation in the 8 to 14 micron infrared band. A large difference
indicates clear skies, whereas a small difference indicates dense,
low-level clouds. This allows the sensor to continuously monitor
the clarity of the skies, and to trigger appropriate alerts on the
control computer. The device also includes a moisture sensor that
directly detects rain drops.

The dome and the operating computer are connected to the mains
power supply through a ``smart'' UPS that will shut down the
observatory in case of an extended power outage.

\begin{figure*}
{\centering
 \includegraphics[clip=,angle=0,width=17cm]{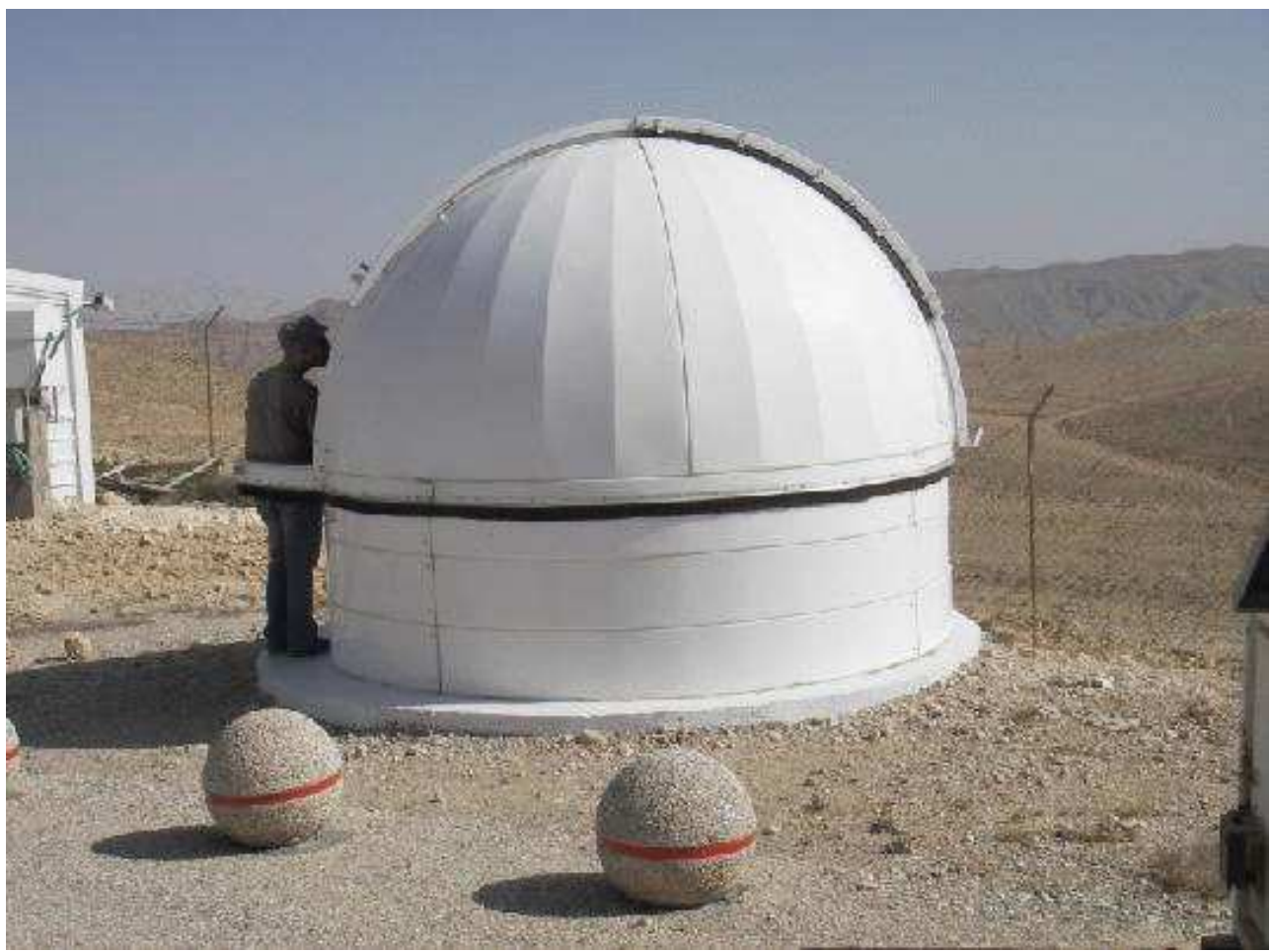}
}
 \caption{%
The C18 dome, with DP looking through the dome shutter to the
telescope.
    \label{fig:C18_dome}}
\end{figure*}

The dome and telescope are mounted on a circular reinforced
concrete slab with a 3.5-m diameter and 0.5-m thickness. The
concrete was poured directly on the bedrock, which forms the
surface ground layer at the WO site. The dome and the concrete
slab are shown in Figure~\ref{fig:C18_dome}.

\section{The operating software}
\label{txt:Software} We decided that all the software would be
tailored into a suite of operating programs that would conform to
the {\it ASCOM} standards\footnote{AStronomy Common Object Model,
http://ascom-standards.org/}. This is in contrast with other
similar, but significantly more expensive, small automatic
observatories (e.g., Akerloff \etal 2003), which chose various
flavors of LINUX. Our choice saved the cost in money and time of
developing specialized software by using off-the-shelf products.
It also facilitated a standard interface to a range of astronomy
equipment including the dome, the C18 mount, the focuser and the
camera, all operating in a Microsoft Windows environment.

The {\it ACP} (Astronomer's Control Program) software is a product
of DC-3 Dreams\footnote{http://acp.dc3.com/}. {\it ACP} controls
the telescope motion and pointing, and can change automatically
between different sky fields according to a nightly observing
plan. {\it ACP} also solves astrometricaly the images collected by
the CCD, and improves automatically the pointing of the telescopes
using these solutions. The program communicates with and controls
the operating software of the dome, interfacing with {\it DDW},
enabling one to open and close the dome shutter, and commanding
the dome to follow the telescope or to go to a ``home'' position.
In addition, the {\it ACP} software is a gateway to the {\it
MaximDL}\footnote{http://www.cyanogen.com/products/maxim\_main.htm}
program that operates the CCD. Different types of exposures,
guiding and cooling of the CCD can be commanded manually using
{\it MaximDL}, but in most cases we use the {\it ACP} code
envelope (the {\it AcquireImages} script) to operate the entire
system. The focusing is also done automatically using the freeware
{\it
FocusMax}\footnote{http://users.bsdwebsolutions.com/~larryweber/},
a software package that operates the Robo-focus and searches
automatically for the best FWHM of a selected star.

\begin{figure*}
\centerline{\includegraphics[angle=0,width=8.5cm]{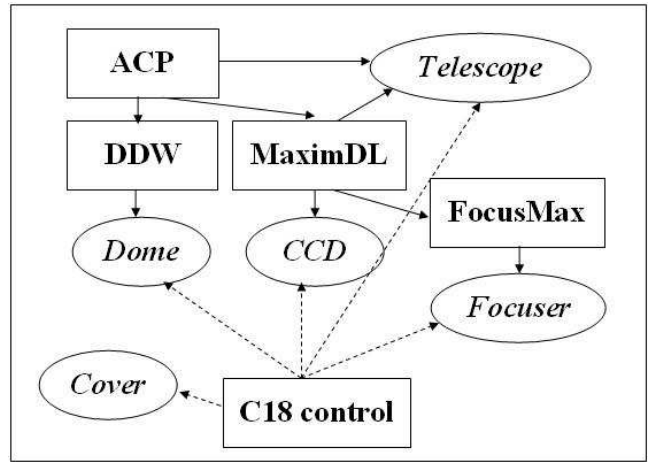}}
 \caption{%
The C18 software hierarchy. Different programs are represented by
rectangles and hardware items by ellipses. Dashed links represent
electrical power on/off connections.
    \label{fig:C18_SW}}
\end{figure*}

To enable power shut-down by remote users we connected the
telescope, dome, CCD, focuser and the telescope's cover to a relay
box that is connected to the six serial ports on the computer.
Using a self-written code ({\it C18 control}) the remote user can
enable or disable the electrical power supply to the different
components of the system and open or close the telescope cover.
This guarantees the safety of the equipment during daytime and
enables the astronomer to fully operate the system from a remote
location using any VNC viewer software. Figure~\ref{fig:C18_SW}
exhibits the different programs making up the software
environment, their connections and their hierarchy.

\section{Cost} The affordability of an automatic telescope is an
important consideration for many observatories. We found that it
was not possible to collect sufficient funds to cover the high
cost of an off-the-shelf automatic telescope with similar
capabilities to those of the installation described here.

The C18 telescope, CCD, and dome, including all the electronic
add-ons and the software, added up to slightly more than 70k\$. To
this one should add significant in-house contributions; the dome
was received in segments that were erected and bolted on the
concrete slab by WO staff Ezra Mash'al and Sammy Ben-Guigui
together with NB; the installation, tuning-up and interfacing of
the various software components were done mainly by IM, DP, and
AB; the polar alignment of the C18 was done (twice) by AB who also
developed the electrically-operated primary mirror cover. All
these and more would add up to about a person-year of work by very
experienced personnel.

\section{Performance}
The automated operation mode of the C18 makes it an easy telescope
to use, demanding from the astronomer 
only to watch the weather conditions when these are likely to
change. Overall, the C18's performance and output are satisfying.

The telescope slewing time is about 5$^{\circ}$ sec$^{-1}$ on both
axes and the settling-down time is 5-8 sec. The dome rotates at a
rate of 3.33 degrees per second in each direction, while its
response time in following the telescope motion is 2-3 sec.
Opening the dome shutter requires 55 sec and closing it 70 sec,
with the dome parked at its ``home'' location. Opening or closing
the electrical telescope cover requires 20 sec.

The auto-guiding system, which uses a smaller CCD in the same
camera head, maintains round stellar images even for the longest
exposures 
that are limited by the sky background, although the telescope is
not perfectly aligned to the North. Without the guider, one can
expect round star images only for exposure times of 90 sec or
shorter. In some cases, telescope shake is experienced due to wind
blows. Since the wind at the WO usually blows from the North-West,
and usually slows down a few hours into the night, selecting
targets away from this direction for the first half of the night
decreases the number of smeared images to a minimum.

The pointing errors of the C18 are of  order 10 seconds of time in
RA and 30 to 60 arcsec in DEC. However, astrometrically solving
the images using the {\it
PinPoint}\footnote{http://pinpoint.dc3.com/} engine, which can be
operated automatically by the {\it ACP} software immediately
following the image readout, re-points the telescope to a more
accurate position for the following images of the same field.

\begin{figure*}
\centerline{\includegraphics[angle=-90,width=8.5cm]{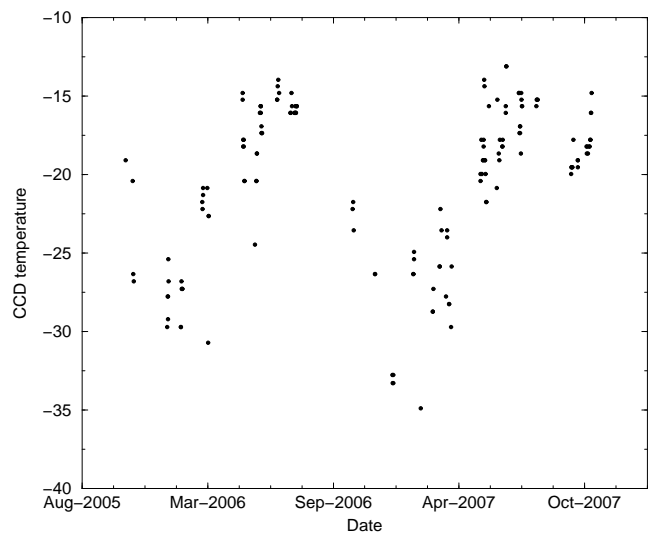}}
 \caption{%
CCD night-time temperature  during two years 
starting on 25 October 2005, as read from the
automatically-generated FITS header of the images. The chip is
warmer in the summer because the hot end of the thermoelectric
cooler is in contact with water at ambient temperature, and this
is warmer in the summer.
    \label{fig:CCDtemp}}
\end{figure*}

Since the CCD cooling is thermoelectric with water-assistance,
with the cooling water at ambient temperature, the CCD temperature
dependents very much on the weather. The usual values run between
-15$^{\circ}$C in the summer nights with ambient temperatures of
+30$^{\circ}$C to -30$^{\circ}$C in the winter (-5$^{\circ}$C
ambient; see Figure~\ref{fig:CCDtemp}). All images are acquired
with the CCD cooling-power at less then 100\% capacity, assuring a
steady chip temperature throughout the observation.

The image read-out using MaximDL takes 15 sec. In regular
observations an additional 10 sec interval is required for the
astrometric solution of the image and another 5 sec to write the
image on the local computer. A five-sec delay is required by the
CCD before it continues to the next image for guider activation.
Since the performance of the astrometric solution and the
activation of the auto-guiding operation are user-selectable, the
off-target time between sequential images is 15--30 sec, yielding
a duty
fraction of 83--92\% on-sky time for typical 180 sec exposures. 

\begin{figure*}[tbh]
\centerline{\includegraphics[angle=0,width=8.5cm]{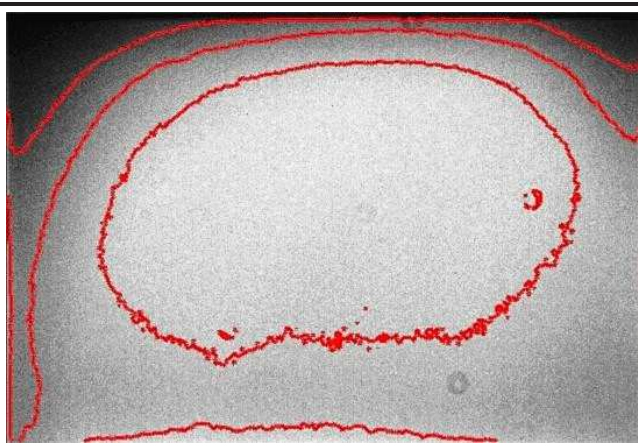}}
 \caption{%
Flat field obtained with the C18 telescope and the SBIG ST-10 XME
CCD. The overplotted contours are spaced by 0.0137\% of the mean
image value, to emphasize the flat field uniformity.
    \label{fig:CCD_FF}}
\end{figure*}

The CCD bias values are 990$\pm$5 counts, a value which is
cooling-dependent. The bias values changed from an original
$\sim$110 to the present value after one year of use and a rebuild
of the software following a disk crash. The CCD flat field (FF)
shows a 1.2\% standard deviation from the image median. The FF
gradient is steeper at the southern corners of the images than at
the northern corners, as seen in Figure~\ref{fig:CCD_FF}. This
probably reflects some additional vignetting in the telescope on
top of the prime focus baffle, possibly caused by the asymmetric
blockage by the SBIG CCD, and by the pick-off prism edge for the
TC-237H tracking CCD, mounted next to the science CCD, which is
used for guiding. Twilight flats give the best results, provided
the CCD cooling is started at least 30 min before obtaining the
flat field images.

On a dark night with photometric conditions the sky background is
10--15 counts per pixel per second. Comparing with the R-band
magnitude of standard stars (Landolt 1992) the sky as measured on
the C18 images in the ``wide-R'' band is about 20.4 mag
arcsec$^{-1}$ and stars of 19.5 magnitudes are detectable with
S/N$\simeq$25. This fits well previous measurements of the sky
brightness at the Wise Observatory ($\mu_R\simeq$21.2 mag
arcsec$^{-1}$ in 1989; Brosch 1992), accounting for the slight
site deterioration in the last 15 years due to ambient light
sources and for the wider spectral bandpass.

\section{Lessons learned}

\subsection{Dome electrical contacts} The dome shutter opening
mechanism receives its 12V DC electrical supply from two spring
contacts that touch metal pads on the dome inner circumference.
These spring contacts are made of slightly elastic copper strips
twisted into rings and are fixed to the dome. To maintain
electrical contact, the copper strips brush against a scouring pad
before reaching the fixed contacts on the dome rim to scrape off
accumulated dirt and oxide.

During the two years of operating the C18 we have had the strips
break a number of times. This interrupts the electrical contacts
to the dome shutter and may prove dangerous when bad weather
arrives in case nobody is present at the WO, since it prevents the
remote shutting of the dome. The remedy is replacing the copper
strips at regular intervals, even though they may appear intact.

\subsection{Dome lift-offs}

We experienced two instance of the dome lifting off its track on
the stationary part of the dome enclosure. In these instances the
dome slips off its track and has to be reseated manually. Since
the fiberglass structure is very light, this operation can be done
by a single person.

We suspect that the dome lift-off events were caused by operator
error. The dome rides on a track that is segmented and passes over
a short door that must be latched shut when rotating the dome. If
this is not done, the door might open slightly causing the dome to
derail.

\subsection{Sticky dome shutter}
We experienced a number of instances when the dome indicated that
the shutter was closed, but inspection with the video camera
showed that the shutter was still open. We have not yet tracked
down the roots of this problem but it may be linked to a
limitation of the electrical power to the shutter motor, perhaps
connected with the operation of a low-voltage DC motor in
conditions of high humidity.

Note that this type of fault requires rapid human intervention to
prevent damage to the telescope and electronics in changing
atmospheric conditions.

\subsection{Operation in a dusty and hot desert environment}

Upon ordering of the PD-10 dome we worried about the possibility
of dust entering the dome while closed, specifically in case of a
dust storm. These storms are rare but can deposit significant
amounts of dust on optics and mechanisms in a short time. This is
why we specified that the dome be equipped with thick brushes that
would prevent most of the dust from entering the (closed) dome.

The two-year operational experience showed that the anti-dust
brushes do indeed protect the interior of the dome and that the
dust deposit on surfaces within the PD-10 dome is similar to that
in the T40 dome. However, after two years of operation, the brush
hairs are no longer straight but show curling and the dust
prevention is no longer optimal. The brushes probably require
replacing.

The daytime external temperature at the WO can reach
+40$^{\circ}$C and the inside of the dome can indeed become very
hot. To prevent this, we plan to air condition the dome during
daytime. This will also cool the huge concrete slab on which the
telescope and dome are mounted, providing a cold reservoir for
night-time operation. Daytime cooling will hopefully improve the
night-time seeing by removing part of the ``dome seeing'' of the
C18.

\subsection{Anti-collision switches}

In remote or automatic operation there exists a possibility that
the telescope may become stuck in tracking mode, bypassing the
software horizon limits. In this case, it is possible that the C18
may drive itself into the concrete floor or into the telescope
fork causing instrumental damage. To prevent this, contact
switches were installed on three corners at the top of the truss
structure. A similar microswitch is activated by the truss
structure if it approaches the base of the telescope fork (see
Figure~\ref{fig:C18_in_dome}). These switches cut off instantly
the power supply through the UPS, in case any of the switches are
activated. Since such an event is primarily an operator error, we
arranged that the reset of the mechanism can only be done
manually, by someone physically pressing the reset button from
within the dome.

\section{Scientific results}

The C18 is used by different researchers at the Tel-Aviv
University for various goals. Some examples, with representative
outputs, are given below. We describe studies of physical
parameters of asteroids,
 the investigation of extrasolar planets, and the monitoring of
variable AGNs.

\subsection{NEOs and other asteroids}
\label{NEOs}

Understanding the potential danger of asteroid collisions on Earth
has encouraged extensive research in recent years. Knowledge of
physical properties of asteroids such as size, density and
structure is critical for any future mitigation plan. At WO we
focus on photometry, which allows the derivation of valuable data
on asteroid properties: periodicity in light curves of asteroids
coincides with their spins; shape is determined by examining the
light curve amplitude; axis orientation is derived by studying
changes in the light curve amplitude; special features in the
light curves, such as eclipses, may suggest binarity and can shed
light on the object structure and density.

The C18 is mainly used for differential photometry of known
asteroids as a primary target, but also enables the detection of
new asteroids. The large field of the CCD allows the asteroids to
cross one CCD field or less per night, even for the fast-moving
Near-Earth Objects (NEOs) that can traverse at angular velocities
of 0.1'' sec$^{-1}$ or slower. This ensures that the same
comparison stars are used while calibrating the differential
photometry. Exposure times between 30 to 180 sec are fixed for
every night, depending on the object's expected magnitude and
angular velocity, the nightly seeing, and the sky background. The
lower limit for the signal to noise ratio of acceptable images is
$\sim$10, thus asteroids that move too fast ($\geq$0.1''/sec) or
are too faint ($\geq$18.5 mag) are de-selected for observation.
Objects brighter than 13 mag are also avoided, to prevent CCD
saturation. The images are reduced using bias, dark and normalized
flat field images. Time is fixed at mid-exposure for each image.
The IRAF {\it phot} function is used for the photometric
measurements. Apertures of four pixel ($\sim$4.5 arcsec) radius
 are usually chosen. The mean sky value is measured in an
annulus with an inner radius of 10 pixels and a width of 10 pixels
around the asteroid. The photometric values are calibrated to a
differential magnitude level using 100-500 local comparison stars
that are also measured on every image of a specific field. For
each image a magnitude shift is calculated, compared to a good
reference image. Stars whose magnitude shift is off by more than
0.02 mag from the mean shift value of the image are removed at a
second calibration iteration. This primarily solves the question
of transient opacity changes, and results in a photometric error
of $\sim$0.01 mag.

Most of the observed asteroids are followed-up on different
nights; this changes the background star field. Some asteroids are
also observed at different phase angles and their brightness can
change dramatically from one session to another. To allow
comparisons and light curve folding to determine the asteroid
spin, the instrumental differential photometric values are
calibrated to standard R-band magnitudes using $\sim$20 stars from
the Landolt equatorial standards (Landolt 1992). These are
observed at air masses between 1.1 to 2.5, while simultaneously
observing the asteroid fields that include the same local
comparison stars used for the relative calibration. Such
observations are done only on photometric nights.

The extinction coefficients and the zero point are obtained using
the Landolt standards after measuring them as described above.
From these, the absolute magnitudes of the local comparison stars
of each field are derived, followed by a calculation of the
magnitude shift between the daily weighted-mean magnitudes and the
catalog magnitudes of the comparison stars. This magnitude shift
is added to the photometric results of the relevant field and
asteroid. The procedure introduces an additional photometric error
of 0.02-0.03 mag. Since the images are obtained in white light,
they are calibrated by the Landolt standards assuming the
measurements are in the Cousins R system. In addition, the
asteroid magnitudes are corrected for light travel time and are
reduced to a Solar System absolute magnitude scale at a 1 a.u.
geocentric and heliocentric distance, to yield H(1,$\alpha^0$)
values (Bowell et al. 1989).

\begin{figure*}[tbh]
\centerline{\includegraphics[angle=0,width=8.5cm]{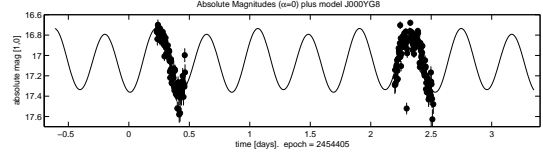}}
 \caption{%
Light curve model fit to the observed photometry data for asteroid
(106836) 2000 YG$_8$.
    \label{fig:LC_fit}}
\end{figure*}

\begin{figure*}[tbh]
\centerline{\includegraphics[angle=0,width=8.5cm]{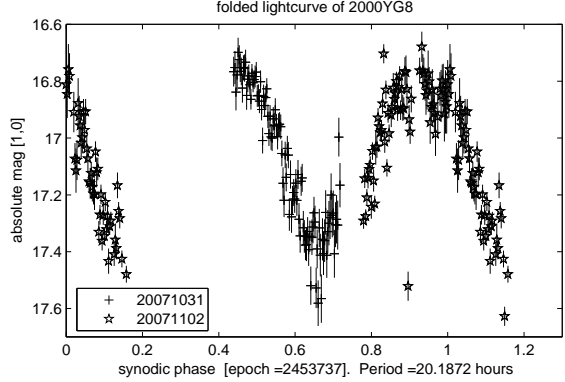}}
 \caption{%
Light curve of the asteroid (106836) 2000 YG$_8$, with the
photometric data folded with a  period of 20.1872 hours.
    \label{fig:LC_folded}}
\end{figure*}

To retrieve the light curve period and amplitude, the data
analysis includes folding all the calibrated magnitudes to one
rotation phase, at zero phase angle, using two basic techniques: a
Fourier decomposition to determine the variability period(s)
(Harris \& Lupishko 1989) and the H-G system for calibrating the
phase angle influence on the magnitude (Bowel et al. 1989). The
best match of the model light curve to the observations is chosen
by least squares. An example is shown on Figure~\ref{fig:LC_fit}
were a simple model was fitted to the observed data points of the
asteroid (106836) 2000 YG$_8$. Figure~\ref{fig:LC_folded} displays
the folded light curve from which the rotation period P is deduced
(here P=20.2$\pm$0.2 hours).

\begin{figure*}[tbh]
{\centering
 \includegraphics[clip=,angle=0,width=17cm]{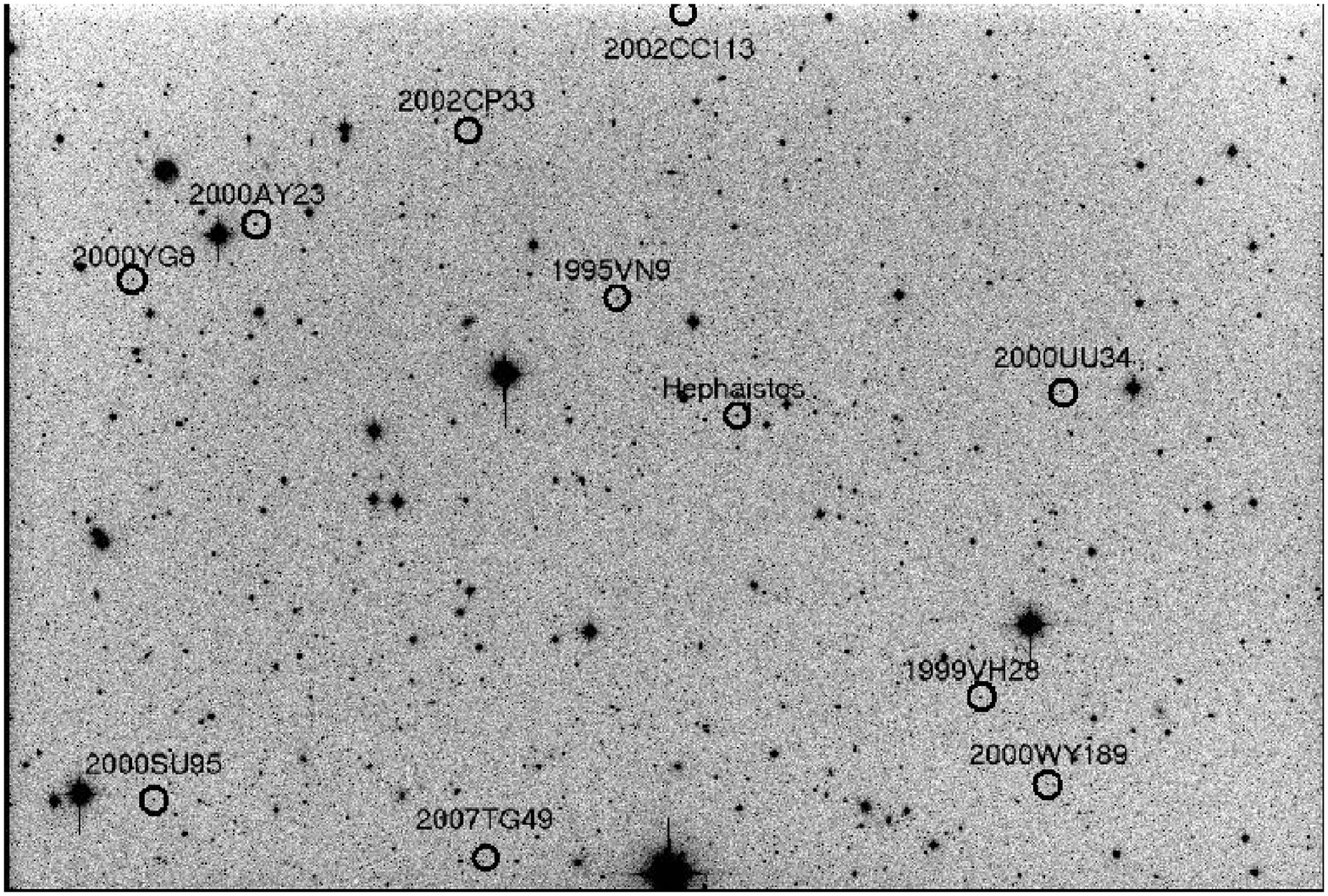}
}
 \caption{%
Eleven asteroids appear on this wide field (40.5'x27.3') image
obtained with the C18 (displayed here as negative). Each asteroid
is labelled with its name and is marked by a circle. The field was
followed for 7.75 hours and the light curve of each asteroid was
measured. The asteroid (106836) 2000 YG$_8$ appears in the
upper-left corner.
    \label{fig:11asteroids}}
\end{figure*}

While the main advantage of the wide field of view of the C18 is
the ability to observe even a fast-moving NEO in the same field
during one night (Polishook \& Brosch 2007), the instrument allows
also the simultaneous monitoring of the light variations from
several asteroids in the same field of view. Looking at Main Belt
asteroids, many objects can be seen sharing the same field (our
recent record is 11 objects in one field; see
Figure~\ref{fig:11asteroids}). An exposure time of $\sim$180
seconds is needed to detect this amount of asteroids while
avoiding the smearing of their images due to their angular motion.

In addition to the rapid increase of our asteroid light curve
database, due to this efficient observing method, new asteroids
are discovered on the same images used for light curve derivation,
proving that 18-inch wide-field telescopes can contribute to the
detection of unknown asteroids even in the age of Pan-STARRS and
other big, automated NEO-survey observatories. We emphasize that
the discovery of new asteroids is {\bf not} a goal of our research
programs, but is a side-benefit. Our records indicate that by
observing in the direction of the main belt we can detect a new
asteroid in every 2-3 fields. These objects are reported to the
MPC, but are not routinely followed up.

The {\it ACP} software can move the telescope automatically
between different fields, increasing the number of measured
asteroids. With the telescope switching back and forth between two
fields, the photometric cost is the increase of the statistical
error by $\sqrt{2}$, since the exposure time is reduced by half
for each field, but the number of measured light curves obtained
every night is very high.

\subsection{Extrasolar planets}

A \emph{transiting} extrasolar planet crosses the parent star's
line-of-sight once every orbital revolution. During this crossing,
referred to as a transit, which usually lasts a few hours, the
planet blocks part of the light coming from the stellar disk,
inducing a $\sim$1\% decrease in the star's observed intensity. By
combining the photometric measurement of the transit light curve
with a spectroscopic measurement of the planet's orbit, both the
planet radius $r_p$ and mass $m_p$ can be derived. Such an
intrinsic planetary characterization can only be done for
transiting planets. By comparing the measured $r_p$ and $m_p$ with
planetary models (e.g., Guillot et al.\ 2005, Fortney et al.\
2007) the planet's structure and composition can be inferred.
This, in turn, can be used to test predictions of planetary
formation and evolution theories (Pollack et al.\ 1996, Boss
1997). In addition, transiting planets allow the study of their
atmospheres (Charbonneau et al.\ 2007 and references therein), the
measurement of the alignment, or lack of it, between the stellar
spin and planetary orbital angular momentum (e.g., Winn et al.\
2005) and the search for a second planet in the system (Agol et
al.\ 2005, Holman \& Murray 2005). For the reasons detailed above
it is clear that transiting planets are an important tool for
extending our understanding of the planet phenomenon, hence the
importance of searching for them.

The combination of the light collecting area, large FOV (0.31
deg$^2$) and short read-out time, makes the C18 a useful tool for
obtaining high-quality transit light curves for relatively bright
stars. The large FOV is especially important, since it allows one
to observe many comparison stars, similar in brightness and color
to the target, which is crucial for the accurate calibration of
the target's brightness. Currently the C18 is used for obtaining
transit light curves for several projects, two of which are:

\begin{itemize}

\item \emph{Photometric follow-up of transiting planet
candidates}: Over the last few years small-aperture wide-field
ground-based telescopes have been used to detect $\sim 15$
transiting planets\footnote{For an updated list of known
transiting planets see:
http://obswww.unige.ch/$\sim$pont/TRANSITS.htm}. This is about
half of all known such planets and the discovery rate of these
instruments is increasing. Once a transit-like light curve is
identified by these small telescopes it is listed as a transiting
planet \emph{candidate}, to be followed-up photometrically and
spectroscopically, discriminating between true planets and false
positives (e.g., O'Donovan et al.\ 2007), and measuring the
system's  parameters. Photometric follow-up observations to obtain
a high-quality transit light curve are carried out to verify the
detection and measure several system parameters, including the
planet radius and mid-transit time. Spectroscopic follow-up is
used for measuring the companion's orbit from which its mass is
inferred. The C18 is used for photometric following-up of
candidates identified in data obtained by the WHAT telescope
(Shporer et al.\ 2007), in collaboration with the HATNet
telescopes (Bakos et al.\ 2006).

\item \emph{Photometric follow-up of planets discovered
spectroscopically}: Most of the $\sim 270$ extrasolar planets
known today were discovered spectroscopically, i.e., through the
radial velocity (RV) modulation of the star induced by the planet,
as the two bodies orbit the joint center of mass. The probability
that such a planet will show transits is $\frac{r_s +
r_p}{d_{tr}}$, where $r_s$ and $r_p$ are the stellar and planetary
radii, respectively, and $d_{tr}$ is the planet-star distance at
the predicted transit time. This probability is about 1:10 for
planets with close-in orbits, with orbital periods of several
days. Interestingly, the first extrasolar planet for which
transits were observed, HD209458 (Charbonneau et al.\ 2000, Henry
et al.\ 2000) was the tenth close-in planet discovered
spectroscopically (Mazeh et al.\ 2000). The C18 is taking part in
follow-up observations of planets detected spectroscopically,
especially those newly discovered, in order to check whether they
show a transit signal. The remote operation makes it possible to
carry out observations on very short notice. Observations made
with the C18 were part of the discovery of the transiting nature
of Gls 436 b (Gillon et al.\ 2007), a Neptune-mass planet orbiting
an M dwarf (Butler et al.\ 2004, Maness et al.\ 2007).


\end{itemize}

During an observation of a transit, the PSF is monitored and the
exposure time is adjusted from time to time, keeping the target
count level from changing significantly. A guide star is usually
used. After bias, dark and flat-field corrections, images are
processed with the IRAF/phot task, using a few trial aperture
radii. The target light curve is calibrated using a few dozen
low-RMS stars similar in brightness to the target. As a final
step, the out-of-transit measurements are fitted to several
parameters, such as airmass, HJD and PSF FWHM, and all
measurements are divided by the fit.

\begin{figure}[tbh]
\centerline{\includegraphics[angle=0,width=8.5cm]{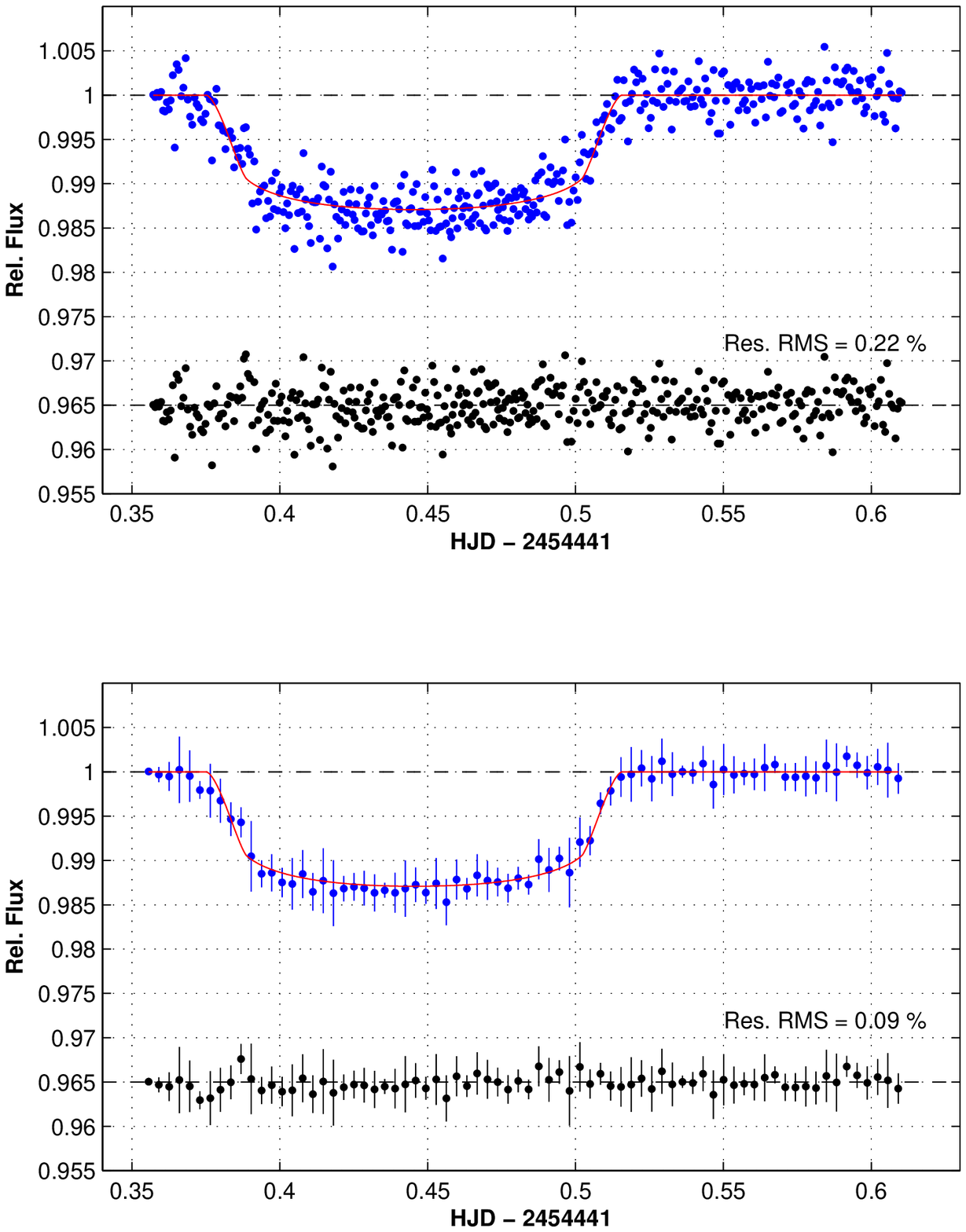}}
 \caption{Light curve of a HATNet transiting
planet candidate obtained by the C18. Both panels show relative
flux vs.\ time. The top panel presents the actual measurements
(blue points) overplotted by a fitted model (solid red line).
Residuals from the fitted model are also plotted (black points),
shifted to a zero point of 0.965. In the bottom panel, blue points
and error bars represent a 5 min mean and RMS. The same model is
overplotted (solid red line) and residuals (in black) are also
shifted to a zero point of 0.965. Each of the 5 min bins includes
5.3 measurements on average. The RMS of the residuals of the
actual measurements is 0.22 \%, or 2.4 milli-magnitude, and for
the binned light curve, 0.09 \%, or 1.0 milli-magnitude. This
light curve includes 400 individual measurements taken during 6.2
hours, with an exposure time of $\sim 25$ sec for each frame.}
\label{fig:trlc}
\end{figure}

\begin{figure}[tbh]
\centerline{\includegraphics[angle=0,width=8.5cm]{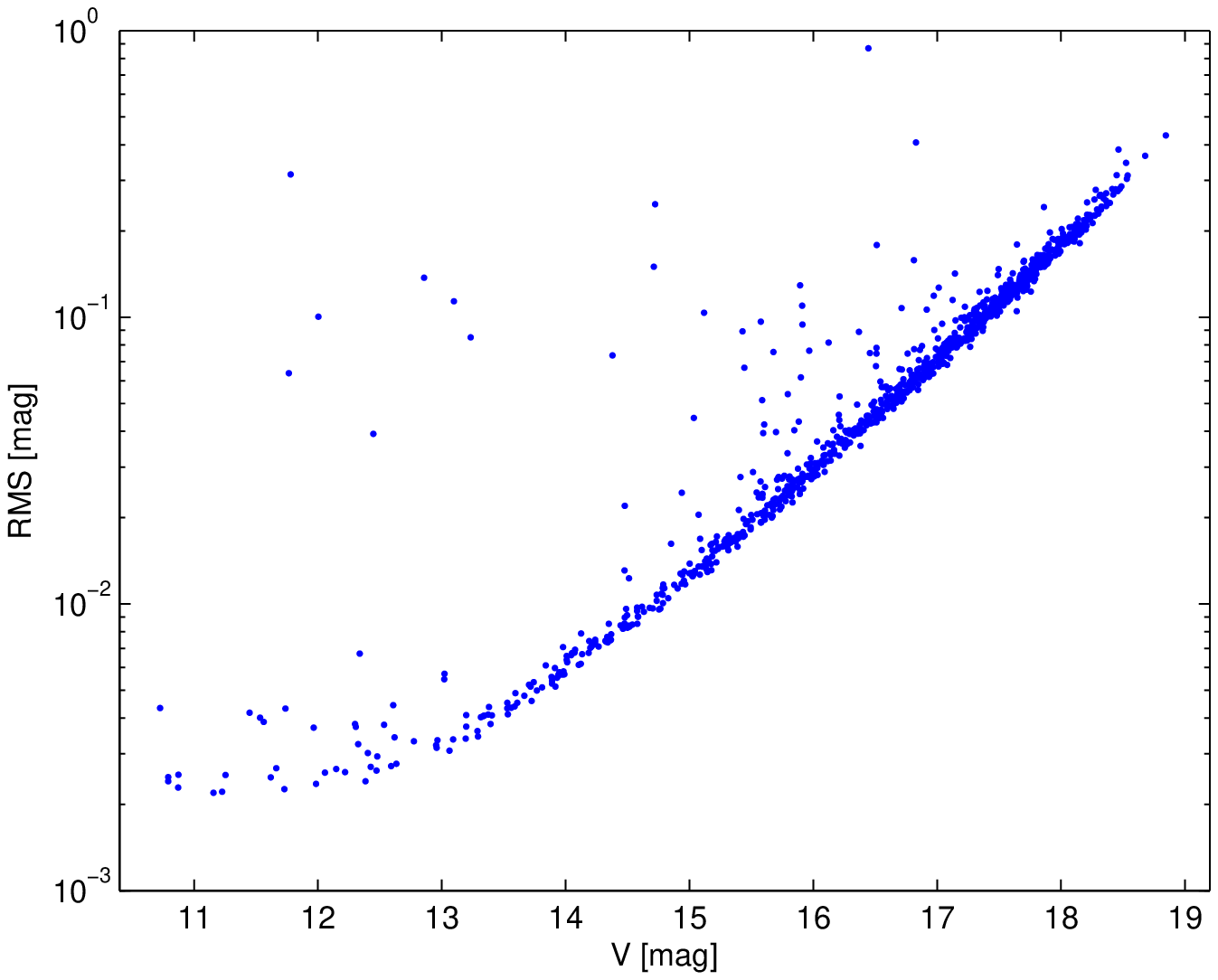}}
 \caption{RMS vs.\ mean $V$ magnitude
for all stars identified in the field of the transiting planet
candidate shown in Fig.~\ref{fig:trlc}. Y-axis is in log scale and
X-axis is in instrumental magnitude, although this is close to the
real magnitude. Observations were done on a single night with a
typical exposure time of 25 seconds. For the brightest stars, the
RMS reaches below the 3 milli-magnitude level.}
\label{fig:rmsmean}
\end{figure}

An example of a transit light curve obtained with the C18 is shown
in Figure~\ref{fig:trlc}; this is the light curve of a HATNet
candidate (Internal ID HTR176-003). The top panel shows the actual
light curve and the bottom panel shows the light curve binned in 5
minute bins. The residual RMS of the unbinned measurements is 0.22
\%, or 2.4 milli-magnitude, and for the binned light curve it is
0.09 \%, or 1.0 milli-magnitude. Figure~\ref{fig:rmsmean} presents
the RMS vs.\ mean $V$ magnitude for all the stars identified in
this field. The X-axis is in instrumental magnitude, which is
close to the real magnitude. The RMS of the brightest stars
reaches below the 3 milli-magnitude level (see also Winn \etal
2007).


\newpage

\subsection{AGN monitoring}

The C18 fits the requirements for monitoring Active Galactic
Nuclei (AGNs). Such objects are known for their continuum
variability and, though the origin of this variability is still
unclear, it is possible to use it to study, using various
techniques, the physical conditions in the AGNs and their
properties. One such technique, which enables the study of the
emission-line gas and the measuring of the central supermassive
black hole mass in AGNs, is the ``reverberation mapping'' (e.g.,
Peterson et al. 1993). In such studies, the time lag between the
variations in the continuum flux and the emission-line fluxes is
used to estimate the Broad Line Region (BLR) size, and to map its
geometry.

Combining such time lags with the BLR velocity (measured from the
width of the emission-lines) allows the determination of the black
hole mass in an AGN (e.g., Kaspi et al. 2000, Peterson et al.
2004). The WO took a leading role in such studies, carrying out
about half of all reverberation mapping studies ever done.

An important empirical size-mass-luminosity relation, spanning a
broad luminosity range, has been derived for AGNs based on 36 AGNs
with reverberation mapping data. This is now widely used to
determine the black hole mass from single-epoch spectra for large
samples of AGNs and distant quasars, allowing the study of black
hole growth and its effect on galaxy evolution. However, more
reverberation mapping studies are needed to expand the luminosity
range and to re-define the current relation with better statistics
(i.e., adding more objects to the 36 already studied).

One crucial element in reverberation mapping studies is the
continuous monitoring of the continuum flux variations. For
low-luminosity AGNs, which vary on timescales of hours to days, it
is important that the monitoring period of several days will be
densely covered at a sampling rate of several minutes. To achieve
such coverage the close collaboration of several observatories
around the world is essential.

\begin{figure}[tbh]
\centerline{\includegraphics[angle=0,width=8.5cm]{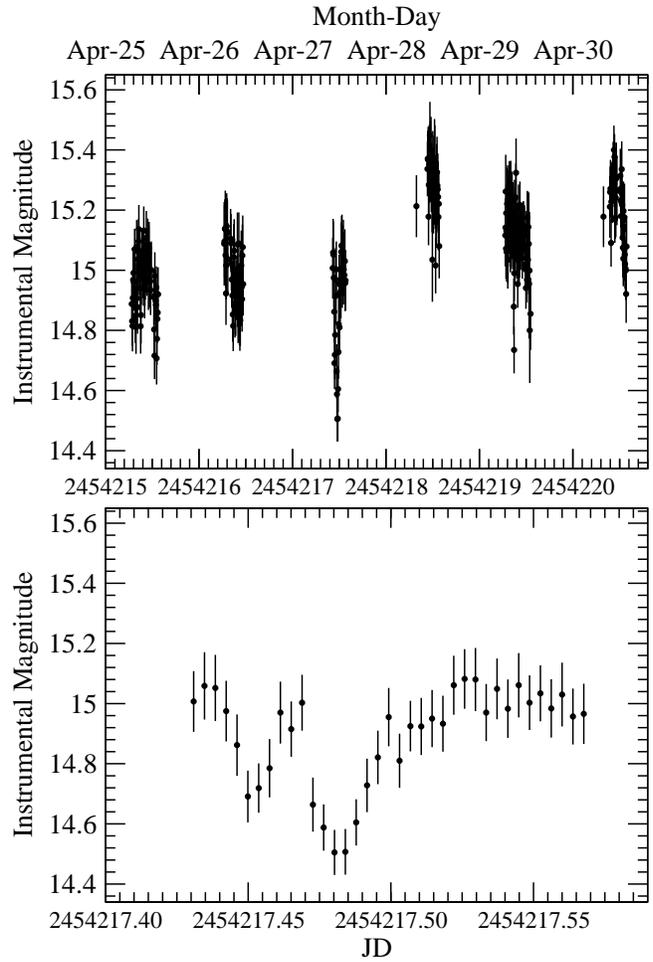}}
\caption{Light curve of the low-luminosity AGN SDSS
J143450.62+033842.5. The full-resolution light curve of the entire
monitoring run is shown in the top panel and a zoomed-in view of
the third monitoring night is shown in the bottom panel. Flux
variations on timescales of hours to days are clearly detected.}
\label{AGN-LC}
\end{figure}

One such  study, \, carried  out with the C18, \, is the monitoring of the
\, low-mass candidate AGN \, SDSS J143450.62+033842.5 (Greene \& Ho
2004; $z=0.0286$, $m_r=15$). During a one-week period in April
2007 this AGN was monitored with a typical exposure time of 240
sec. The images were reduced in the same way as described in
Section~\ref{NEOs}. The typical uncertainty of the measurements is
about 0.1 magnitude which is sufficient to detect the variability
of this AGN. Figure~\ref{AGN-LC} shows the light curve, derived
using differential photometry and the ``daostat'' program
described in Netzer et al.  (1996, section 2.3).

Combining this light curve with additional data from the Crimean
Astrophysical Observatory in Ukraine, the MAGNUM-2m telescope in
Hawaii, and the 2m Faulkes Telescope in Australia, produced a
fully-sampled continuum light curve for SDSS J143450.62+033842.5
covering several days. Together with an emission-line light curve
obtained at the Magellan telescope, the central mass of the black
hole in the center of this AGN will be determined. 
The C18 based at the WO played a crucial part in
this monitoring campaign and proved that it can be most useful in
producing AGN light curves, which are used in state-of-the-art
studies of supermassive black holes.

\section{Conclusions}
\label{txt:Conclusions} We described the construction of a
secondary observing facility at the Wise Observatory, consisting
of a 0.46m Prime Focus f/2.8 telescope equipped with a good
quality commercial CCD camera. The C18 telescope and camera are
sited in a small fiberglass dome, with a suite of programs using
ASCOM interface capability orchestrating the automatic operation.
The combination offers a cost-effective way of achieving a limited
goal: the derivation of high-quality time-domain sampling of
various astronomical sources.

The automatic operation, despite the few snags discovered
following two years of operation, proves to be very user-friendly
since it allows the collection of many observations without
requiring the presence of the observer at the telescope or even at
the Wise Observatory. This was achieved with a very modest
financial investment and may be an example for other astronomical
observatories to follow.

We presented results from three front-line scientific projects
performed with the new facility, which emphasize the value of an
automated and efficient, relatively wide-field, small telescope.

\section*{Acknowledgments}
The C18 telescope and most of its equipment were acquired with a
grant from the Israel Space Agency (ISA) to operate a Near-Earth
Asteroid Knowledge Center at Tel Aviv University. DP acknowledges
an Ilan Ramon doctoral scholarship from ISA to study asteroids. We
are grateful to the Wise Observatory Technical Manager Mr. Ezra
Mash'al, and the Site Manager Mr. Sammy Ben-Guigui, for their
dedicated contribution in erecting the C18 facility and repairing
the many small faults discovered during the first two years of
operation.

\end{document}